# MODELLING THE SPREAD OF COVID-19 ON MALAYSIAN CONTACT NETWORKS FOR PRACTICAL REOPENING STRATEGIES IN AN INSTITUTIONAL SETTING.

Fatimah Abdul Razak[1*] & Paul Expert[2,3]

[1]*Department of Mathematical Science, Faculty of Science and Technology, Universiti Kebangsaan Malaysia, 43600 UKM Bangi, Selangor, Malaysia*
[2]*Global Digital Health Unit, Imperial College London, SW7 2AZ, U.K.*
[3]*Tokyo Tech World Research Hub Initiative, Tokyo Institute of Technology, Japan*
*Corresponding author: fatima84@ukm.edu.my

**Abstract**

Reopening strategies are crucial to balance efforts of economic revitalization and bringing back a sense of normalcy while mitigating outbreaks and effectively flattening the infection curve. This paper proposes practical reopening, monitoring and testing strategies for institutions to reintroduce physical meetings based on SIR simulations run on a student friendship network collected pre-Covid-19. These serve as benchmarks to assess several testing strategies that can be applied in physical classes. Our simulations show that the best outbreak mitigation results are obtained with full knowledge of contact, but are also robust to non-compliance of students to new social interaction guidelines, simulated by partial knowledge of the interactions. These results are not only applicable to institutions but also for any organization or company wanting to navigate the Covid-19 ravaged world.

**Abstrak**

Strategi pembukaan semula sangat penting dalam menyeimbangkan usaha pemulihan semula ekonomi tanpa mengabaikan usaha melandaikan lengkung jangkitan. Makalah ini mencadangkan beberapa strategi pengawasan dan pengujian yang praktikal bagi mengadakan perjumpaan fizikal berdasarkan simulasi ke atas rangkaian persahabatan pelajar yang dikumpul sebelum penyebaran Covid-19. Simulasi ini boleh digunakan sebagai penanda aras untuk menguji beberapa strategi pembukaan semula yang boleh digunapakai dalam kelas. Simulasi kami menunjukkan bahawa strategi terbaik diperolehi dengan mengetahui maklumat penuh rangkaian perhubungan. Namun begitu, kami juga mencadangkan strategi yang tidak perlukan maklumat rangkaian. Dapatan ini bukan sahaja terhad untuk institusi tetapi ia juga boleh diguna pakai oleh sebarang organisasi atau syarikat dalam menghadapi norma baharu era Covid.

Kata kunci: Covid-19, rangkaian kontak, rangkaian persahabatan, strategi pembukaan semula, SIR dalam rangkaian

**Introduction**

Covid-19 has forced humanity to adapt to new interaction norms while keeping an eye on the number of cases, the growth of which must be flattened to ensure they do not grow exponentially and overwhelm the health services. This curve representing the number of cases is commonly modelled by the Susceptible-Infected-Recovered (SIR) model (Keeling & Eames 2005; Pastor-Satorras et al. 2015) and its variants. More extensive models (Ferguson et al. 2020; Chinazzi et al. 2020) utilize estimated contact networks derived from population level observations as the underlying model of the SIR spread especially in light of the contact-based transmission of Covid-19. Therefore, contact networks are crucial elements to inform realistic and effective reopening strategies in the face of Covid-19.

In this paper, our aim is twofold. Firstly, we shall model the spread of Covid-19 on a real Malaysian contact networks of first year university students. We shall simulate the dynamics of potential super spreaders in an institutional setting. Second, we shall propose practical reopening strategies for institutions based on the concept of centralities and friendship paradox on networks.

**Contact Networks**

Human to human interaction is key to the transmission of infectious diseases in general and Covid-19 in particular. An intuitive way to view human interactions is through contact networks (Yang & Jung 2020; Herrera et al. 2016). For example, two people are considered connected if they have been in contact at workplaces, households or schools. Underlying contact networks have been used to model previous outbreaks such as Influenza (Ferguson et al. 2006), Measles (Hunter et al. 2018), the 2019 H1N1 (Pastor-Satorras et al. 2015; Liu et al. 2018) and the MERS in 2015 (Yang & Jung 2020).

A contact network formed through connections of individual and hospitals where individuals are connected by being at the hospital at the same time, highlighted that nosocomial infection was the main cause for the 2015 Middle East Respiratory Syndrome (MERS) outbreak in Korea (Yang & Jung 2020). The structure of contact networks can readily explain the existence of super-spreaders and clusters formation in epidemic spreading (Yang & Jung 2020, Herrera et al. 2016, Rudiger et al. 2020) since these behaviours are common observations on real networks known in network analysis as hubs and communities respectively.

The Global Epidemic and Mobility Model (GLEAM), a meta-population based, spatial epidemic model have utilized transport networks on top of contact networks to project the impact of travel limitations on the national and international spread of the epidemic. GLEAM (Chinazzi et al. 2020; Vibouda & Vespignani 2019) highlighted that the travel quarantine of Wuhan delayed spread at the international scale, where case importations were reduced by nearly 80% until mid-February. The effect of social distancing, testing and quarantine efforts were also observed by Aleta et al. (2020).

Contact networks offer an avenue to test the effect of non-pharmaceutical interventions such as closing down workplaces, community centers and schools (Ferguson et al. 2006; Chinazzi et al. 2020, Perra 2021) and various combinations of travel restrictions and partial or complete lockdown. Networks can be utilized to plan strategies for vaccinations, mitigation, suppression (Fu et al. 2017, Hunter et al. 2018; VanderWeele & Christakis 2019) and the design of early warning systems for disease surveillance, especially when coupling physical interactions with virtual interaction through social media (Herrera et al. 2016; VanderWeele & Christakis 2019). Furthermore, once a network is obtained, various techniques of analysis can be utilised to understand how its structure affects the dynamic taking place on it (Zulkepli et al. 2020; Razak & Shahabuddin 2018).

Contact tracing has played a crucial role in the Malaysian response especially in the initial stages of the pandemic. A combination of thorough questioning and isolating all possible contact even before symptoms are manifested can be effective. The contact network underlying the contact tracing information is the reason why the strategy of contact tracing and isolation is highly effective in reducing the number of cases. Malaysian interaction networks have been modelled in (Ratanarajah et al. 2020; Razak et al. 2019).

**University Friendship Networks**

Network analysis is a tool based on graph theory. A network can be written as $G = (V, E)$, where $V$ is the set of vertices and $E$ is the set of edges (Newman, 2010). Network are combinatorial objects used to model relations between elements of a system. Typically, a network is depicted in diagrammatic form as a set of dots for the vertices, joined by lines or curves for the edges as in Figure 1. We define the set of neighbours of a vertex as the vertices connected to it via an edge. Figure 1 is a network with the set of vertices $V$ representing 148 individuals and the set of edges $E$ representing the friendships amongst them as collected in pre-Covid physical class setting through questionnaires (Rahman et al. 2020; Ratanarajah et al. 2020). All participants were fully informed that their data would be used for research. The data sets were anonymised, stored and analysed in a secure environment.

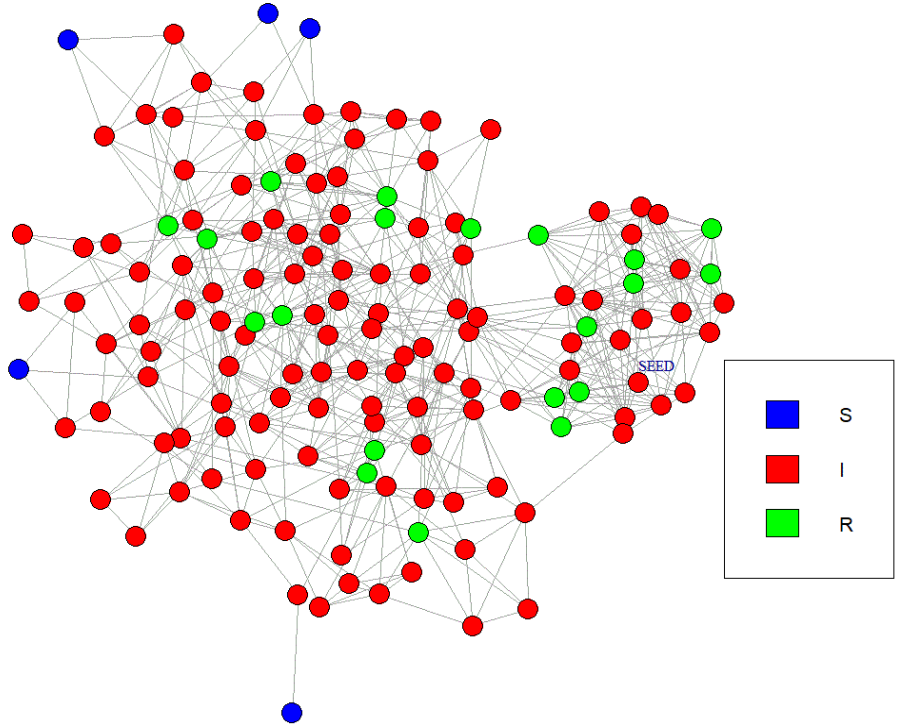

Figure 1. A network of 148 individuals as vertices linked by edges representing friendship relationships. Colours represent the state of each vertices - Susceptible (Blue), Infected (Red) or Recovered (Green) - at the peak infection time t=13 during a simulation run of the SIR model. The seed vertex is labelled. The number of individuals in each state as a function of time are plotted in Figure 2.

**SIR spread on Networks**

The basic SIR model is governed by the equations

$$\frac{dS}{dt} = -\beta SI, \quad \frac{dI}{dt} = \beta SI - \gamma R, \quad \frac{dR}{dt} = \gamma R, \qquad (1)$$

where $\beta$ is the infection rate and $\gamma$ is the recovery rate (Keeling & Eames 2005, Pastor-Satorras et al. 2015). This model assumes that the population is categorized into three groups, namely the 'Susceptible', the 'Infected' and the 'Recovered'. $S$ is the number of individuals being in the 'Susceptible' state, $I$ the number of 'Infected' individuals and $R$ the number of individuals in the 'Recovered' state.

The basic SIR model assumes homogeneity and full mixing, i.e. all individuals interact with all other individuals and with the same probability at all times. Using contact networks adds heterogeneity to the modelling not only by exactly specifying who is in contact with whom but networks can also be customized to account for age, comorbidity, gender, different types and strains of viruses (Forster et al. 2020; Fu et al. 2017) or mutating pathogens that change infection rates (Rudiger et al. 2020). Real networks are very useful to quantify the extent to which real populations depart from the homogeneous-mixing assumption, in terms of both the underlying network structure and the resulting epidemiological dynamics.

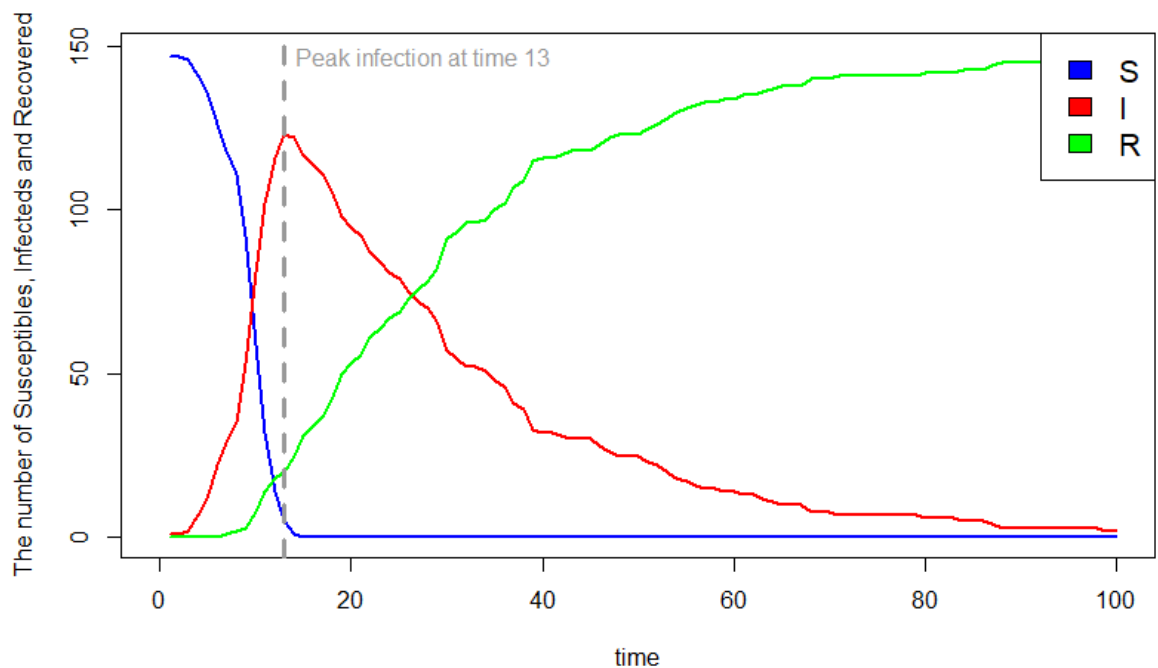

Figure 2. Outcome of a simulation run of the SIR model on the students contact network for 100 time steps. $S$, $I$ and $R$ represent the number of individuals being in the 'Susceptible', 'Infected' and the 'Recovered' state at every time step respectively. In this simulation, even

with one randomly selected initial infected individual, the pathogen can spread through the whole population. No interventions, such as the isolation of infected individuals, were implemented. The grey dotted line represents peak infection time and Figure 1 is the network visualization of the individuals states at this time step.

Figure 2 is the typical outcome of an SIR simulation on the student network where the pathogen spreads through the whole population with no intervention. In this paper we assume that Recovered individuals can no longer be infected. We simulate the SIR process on networks in this paper with $\beta = 0.054$ and $\gamma = 0.05$ in conforming with $R_0 = \frac{\beta}{\gamma} = 1.08$ given by the MySejahtera app as of 25/1/2021 and define the unit of time to be days. We assume that testing can be done on each day for the purposes of monitoring and intervention. MySejahtera is the official app the Government of Malaysia developed to provide statistics of infections and assist in monitoring the Covid19 outbreak.

The final size of the outbreak, quantified by the total number of infections, depends on the ratio between infection and recovery rates: $R_0$. $R_0$ can be calculated using $R_0 = \frac{\beta}{\gamma}$ (Newman 2010) where $\beta$ and $\gamma$ are from equation (1). $R_0$ represents the average number of individuals infected by one infected individual and should not be confused with $R$ representing the number 'Recovered' individuals. This number characterises the virulence of the pathogen modelled.

To model the epidemic spread on networks, an agent-based model approach is usually taken in replicating the SIR spread (Ferguson et al. 2020; Chinazzi et al. 2020). In a network setting, transmission can only occur between vertices that are connected to each other. Therefore, the infection spreading depends on the connectivity of each infected vertex. In this paper we capture the idea of the basic SIR model on a network by setting the model parameters of infection rate $\beta$ and recovery rate $\gamma$ equal for each vertex in the network, assuming homogeneity of the student population response to Covid-19. The main mechanism is summarized in these steps.

1. Infected vertices recover exponentially at rate $\gamma$.
2. Infected vertices may infect susceptible neighbours at an exponential rate $\beta$ until it recovers.
3. Recovered vertices are not infectious and can no longer be infected.

Thus, at each time step every vertex is in one of the *S*, *I* or *R* states. Figure 1 depicts a visualization of our network at peak infection in one of our simulations. In this particular realization, 83% of the population is infected at time 13, which corresponds to peak infection time as seen in Figure 2.

**Flattening the Infection Curve**

The infected curve resulting from the SIR process, the red line in Figure 2, representing the numbers of infected individuals at every time step, is the oft-repeated curve that requires flattening. The SIR process is stochastic, therefore one needs to run several iterations to get a representative outcome when modelled on a heterogeneous interaction medium such as a network, as the location of the seed patient in the network will impact the speed at which the epidemics will spread through the system.

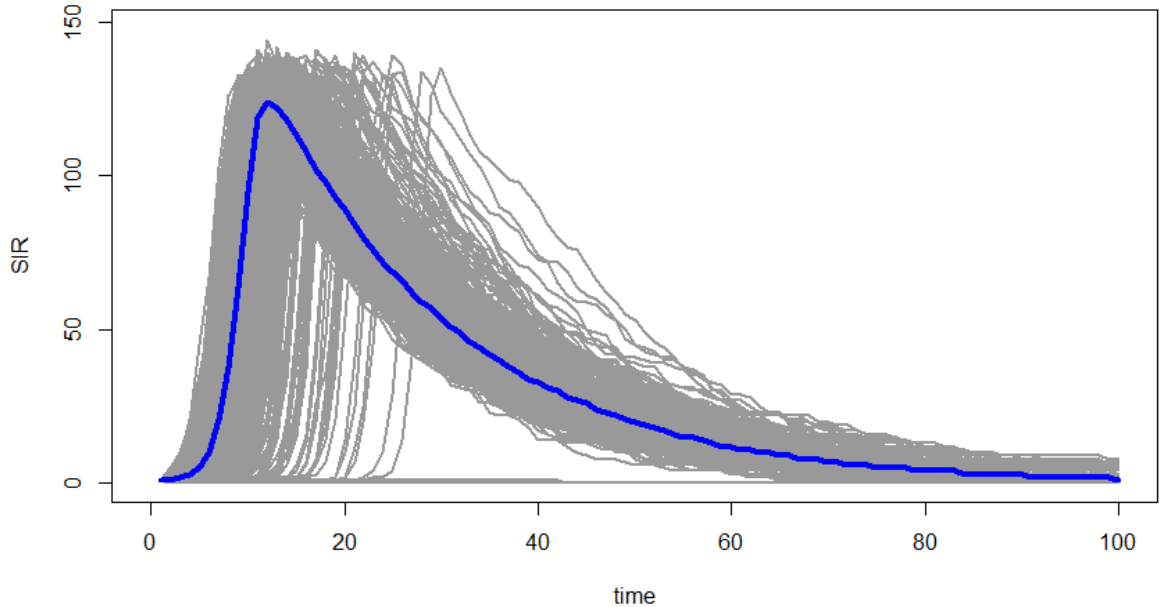

Figure 3. 1000 iterations of the simulated SIR on the network. The Median behaviour highlighted in blue peaks at time 12 with 84% of the population infected.

Therefore, to initialize the infection we randomly choose one single vertex at random for every simulation run, so as to qualify the randomness of the simulation as a reflection of the contact structure rather than the stochastic nature of infection and recovery. The former being the important thing to consider when devising monitoring strategy: you want it to be effective whatever the origin of the infection. In Figure 3, we simulated 1000 iterations for 100 time steps and highlighted the median in blue. As we are interested in the evolution to peak time of an outbreak, we set a maximum simulation time of a 100 time steps, as peak time never exceeds 50 time steps in our simulations.

For each time step, we use the median value over all iterations and take it as the most likely evolution of the process. From our simulations of randomly infecting one vertex at every run, we observed that the median for 1000 and the first 100 iterations both resulted in peaks at time 12 with 84% of the population infected at the peak. Thus in this paper, all the simulations results in Figures 4-9, are represented by the median obtained from 100 runs of 100 time steps with $\beta = 0.054$ and $\gamma = 0.05$ with a randomly chosen initial infection seed for every run replicated using the software R.

The clarion call to physical distancing was instigated by the need to flatten the infection curve by reducing interactions between individual and thus pathogen spreading opportunities, thus reducing the total infected number to a fraction of the population. For example in Figure 3, the flattening can be achieved both by reducing the percentage of maximum infected population from 84% and/or by delaying the time of peak infection. Using a network representation of contacts makes the mechanism by which social distancing operates clear: if all edges are severed, no infection can spread. In reality this effect is achieved by testing and isolating the infected individuals as well as all their contacts regardless of symptoms, effectively breaking infection chains in the network, as the pathogen can only spread along edges.

The student friendship network gives an estimation of the most likely contacts between individuals. By analysing the structure of the network, the monitoring process of the population can be simplified. We simulate the effect of monitoring by "testing" a pre-selected subset of vertices at every time step. If a vertex is tested positive, i.e. it has transitioned from the susceptible to the infected state, it is removed from the network, i.e. quarantining, by effectively having its edges temporarily deleted, until it becomes recovered, thus breaking the transmission chain. The main mechanism is summarized in these steps.

1. Infected vertices recover exponentially at rate $\gamma$.
2. Infected vertices may infect susceptible neighbours at an exponential rate $\beta$ until it recovers unless the vertex is in the monitored group in which case the vertices and all neighbours shall be quarantine.

3. Recovered and Quarantined vertices are not infectious and can no longer be infected.

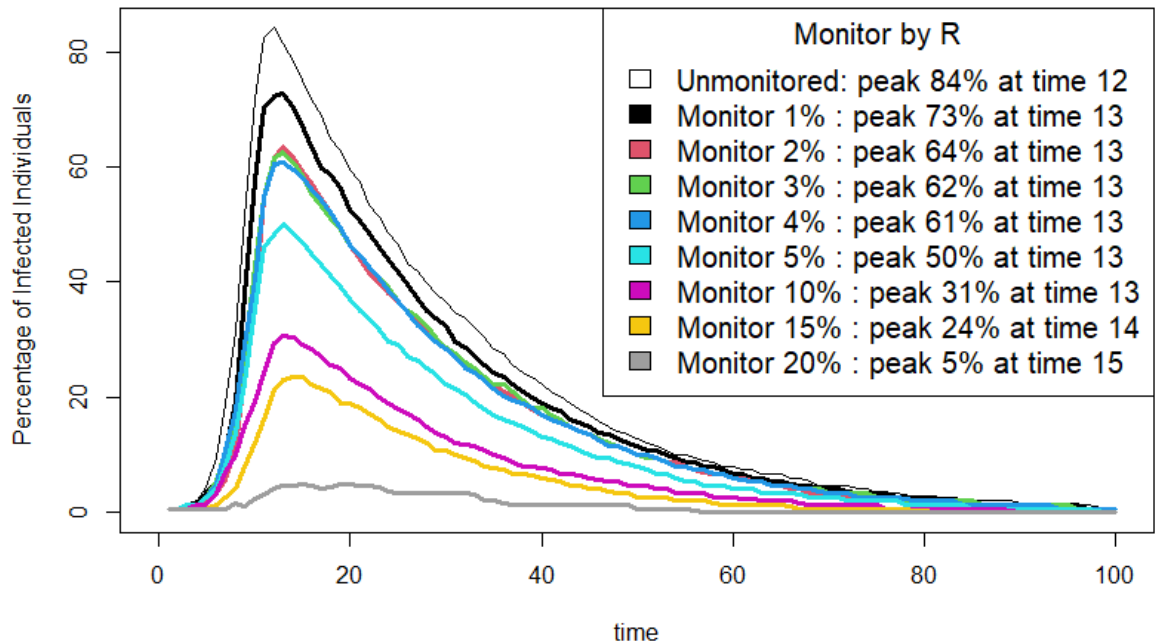

Figure 4. Medians of the 100 iterations simulated for randomly selected groups of students to be monitored. Colours represent different monitored percentage.

Figure 4 illustrates median curves of infection obtained by monitoring a randomly selected percentage of the population. The random selection is done separately for each of the 100 iteration. The black line for example represents 1%, or 2 individuals, monitored out of the 148 individuals. As the number of monitored individuals increase, the peak number of infected individuals decreases from 84% with no monitoring to 5% with 20% of students randomly monitored and the peak time is delayed from 12 to 15.

The question is: can we do better by targeting individuals based on their position in the contact network rather than randomly monitoring a fraction of the population? The aim of all monitoring strategies is to identify vertices in the network whose removal when infected would reduce the spreading as much as possible, or in other words how to reduce the connectivity of a network in the most efficient way to slow down or even suppress the epidemic spreading (Holme 2002; Watanabe & Masuda 2010). It is a similar problem to the planning of optimal immunisation and vaccination strategies (Holme & Litvak 2017; VanderWeele & Christakis 2019; Pastor-Satorras & Vespignani 2002). In this paper, we focus on the role of vertices in the contact network to devise monitoring strategies.

### The importance of individuals in the contact network

Networks, in general, possess a heterogeneous structure, and it is therefore possible to quantify topological differences between nodes. This naturally leads to the idea the importance, or centrality of a vertex in the network with respect to a metric. The concept of centrality originates from the discipline of social network analysis (Freeman 1977; Freeman 1979). Defining a value for each vertex according to a centrality measure implies that vertices can be ranked. A next natural problem is to find which measure of a vertex role correlates the best with observables from a dynamical process. For example, which are the individuals that if monitored for infections status and removed when infected, mitigate the most the number of total infected individual or shift the peak time the most. A similar problem is to identify individuals whose vaccination would reduce the most the spread of an epidemic.

There exists a slew of centrality metrics that each measure particular aspects of a vertex, but we will use the three 'original' ones introduced by Freeman: degree, closeness and betweenness centrality that we define precisely below. Despite their simplicity, they capture particularly well relatively orthogonal and essential properties of vertices that are of interest when considering spreading dynamics: direct connectivity, ease of reach from other vertices and role in bridging subparts of a network. Perhaps the simplest centrality measure is the degree centrality (DC). The degree $k_i$ of a vertex $i$ is the number of edges connected to it: the number of it's direct neighbours. The effective infection rates will be proportional to the number of neighbours a vertex has, therefore it makes sense to monitor individuals with the highest degree (Holme and Litvak 2017). In Figure 4, the monitored group was randomly chosen but in Figure 5 the monitored group consist of individuals with the highest DC. In order to rank by DC, we first calculate the degree $k_i$ of each vertex $i$. We then rank the vertices according to $k_i$ using the sort function in software R. This needs to be done only once for all the simulations because the networks is static. The black line in Figure 5 represents 1%, or 2 individuals with the highest DC, monitored out of the 148 individuals. The grey line in Figure 5 represents 20%, which entails the monitoring 30 individuals with the highest value of $k_i$. Figure 5 illustrates the reduction in the number of infected individuals when a certain percentage of individuals with the highest DC is monitored.

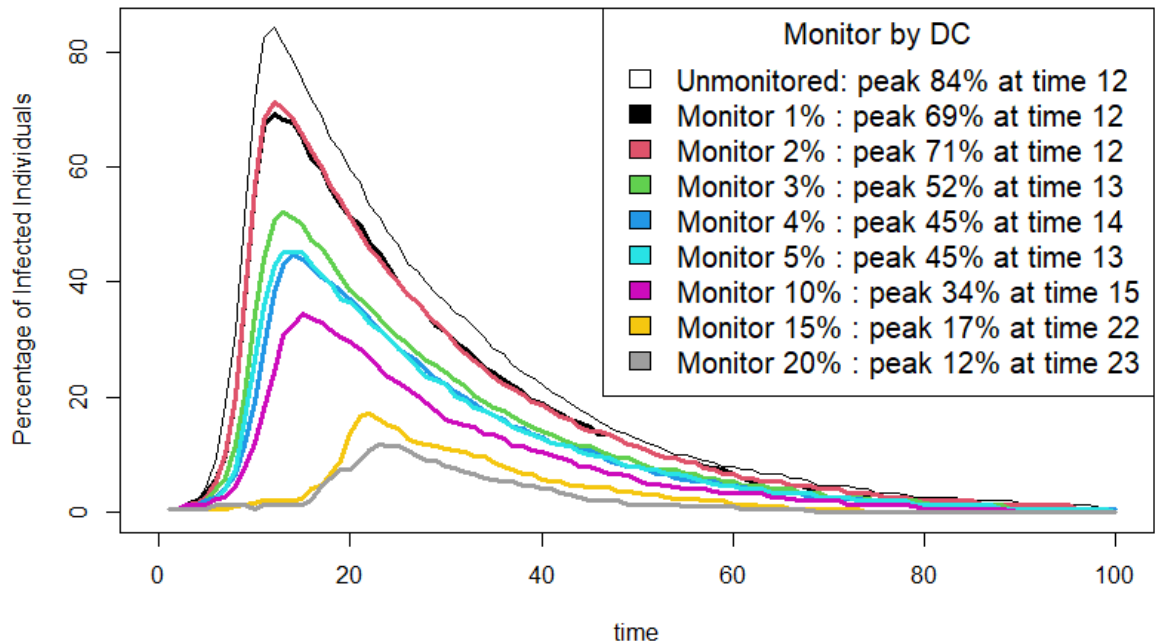

Figure 5. Medians of the 100 iterations simulated for groups selected from those with the highest degree $k_i$ to be monitored. Colours represent different monitored percentage.

A path in a network is any sequence of vertices such that every consecutive pair of vertices in the sequence is connected by an edge in the network. The distance $d_{ij}$ between two vertices, $i, j \in V$ is defined as the number of edges along the shortest path, i.e. minimum number of edges used to connect them. Every pair of vertices directly connected by an edge are thus at a distance of 1. The Closeness Centrality (CC) of vertex $i \in V$ can be defined as the reciprocal of the average distance a vertex is from all other vertices (Newman 2010; Freeman 1979), such that

$$l_i = \frac{1}{\sum_{j \in V} d_{ij}} \qquad (2)$$

It uses the inverse sum of distances from a vertex to all other vertices in the network to rank the vertices. The larger the CC value the 'closer' the vertex is to all other vertices. Therefore it makes sense to monitor the vertices with highest CC as illustrated in Figure 6 where 20% monitoring renders almost no other infections other than the initial one forced onto the simulation. In order to monitor by CC, we used the same steps utilized for DC.

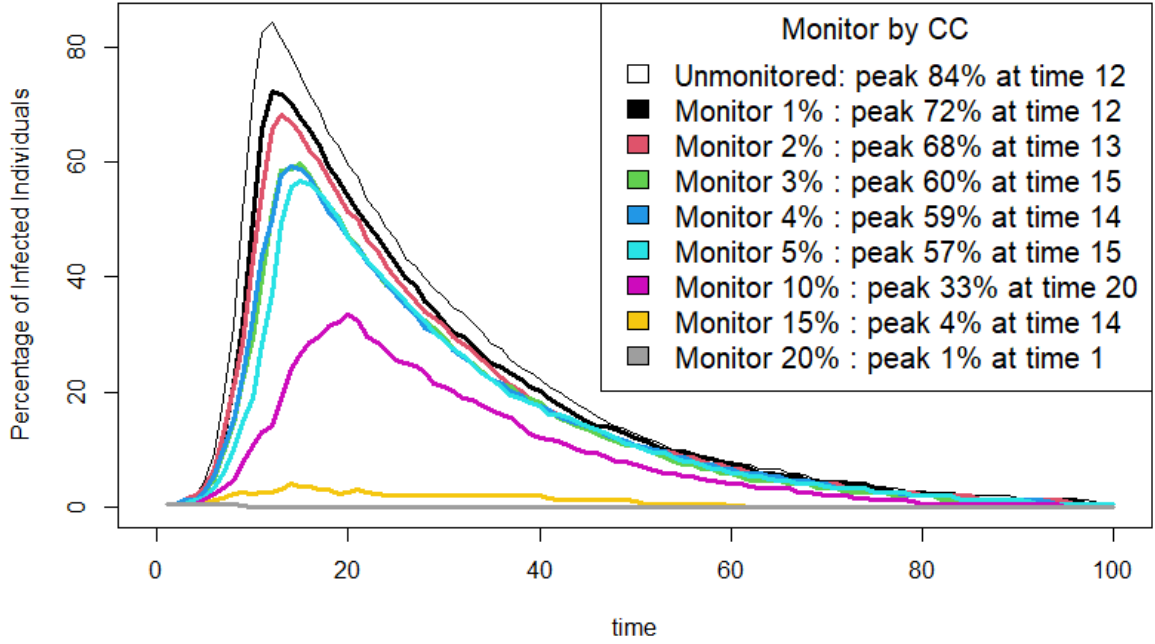

Figure 6. Medians of the 100 iterations simulated for groups selected from those with the highest Closeness Centrality $l_i$ to be monitored. Colours represent different monitored percentage.

The Betweenness Centrality (BC) of vertex $i \in V$ can be defined (Freeman 1977; Newman 2010) as

$$x_i = \sum_{s \neq i, s \in V} \sum_{t \neq i, t \in V} \frac{n_{st}(i)}{N_{st}} \qquad (3)$$

where $n_{st}(i)$ is the number of shortest paths from $s$ to $t$ that passes through vertex $i$. $N_{st}$ is the total number of shortest paths from $s$ to $t$. The more "in-between" other vertices a vertex is the more central it is. Clusters of vertices that are more connected among themselves that with others are a common feature of complex networks (Girvan & Newman 2002). High BC vertices can act as connectors between different clusters formed in the network. Thus by quarantining these vertices the separate parts of the network become less connected with each other therefore a larger proportion of the network can be protected from infection as evidence in Figure 7 where BC monitoring performs better than all the other groups. In order to monitor by BC, we used the same steps utilized for DC.

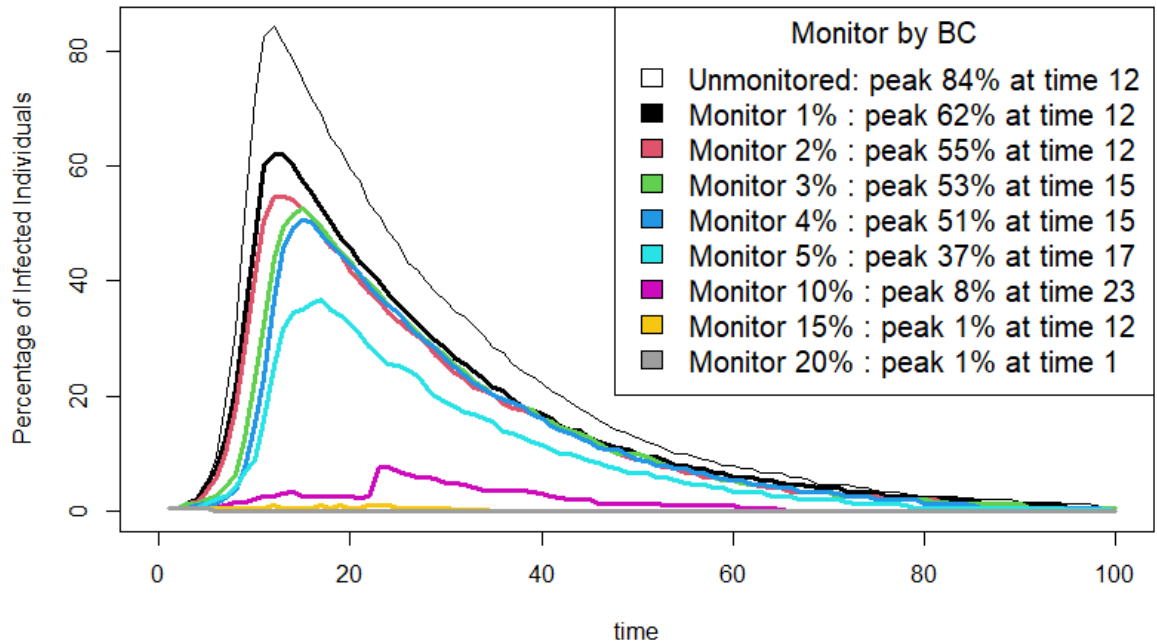

Figure 7. Medians of the 100 iterations simulated for groups selected from those with the highest Betweenness Centrality $x_i$ to be monitored. Colours represent different monitored percentage.

**Reopening Strategies and the Friendship Paradox**

An ideal reopening strategy would be to know the actual contact networks of all individuals in the institution in real time thus providing full knowledge of the network and consequently enabling monitoring of high centrality individuals to prevent super spreader events. However gathering such information about contact networks may not be practical in many situations, particularly contact tracing apps that are typically run by government rather than individual institutions that only have more limited resources available. Questionnaire based contact networks also have their limitations unreported non-compliance to new social interaction guidelines by having contact outside prescribed "bubbles", render the underlying contact network incomplete. In these cases we only have partial knowledge of the network.

The friendship paradox is the phenomenon whereby most people have fewer friends than their friends have. Specifically that the average number of friends of friends is always greater than the average number of friends of individuals. In a friendship network, the numbers of friends of vertex $i$ is represented by the degree $k_i$ thus the average number of friends of individuals is $\frac{\sum_{i=0}^{N} k_i}{N}$ where $N$ is the size the network. Whereas, the average number of friends of friends can be obtained by $\frac{\sum_{i=0}^{N} k_i^2}{\sum_{i=0}^{N} k_i}$. Therefore the friendship paradox on a network (Feld 1991) can be expressed by

$$\frac{\sum_{i=0}^{N} k_i^2}{\sum_{i=0}^{N} k_i} \geq \frac{\sum_{i=0}^{N} k_i}{N} \tag{4}$$

This is true for all networks. Particularly, for our student friendship network with $N = 148$, we obtain that $\frac{\sum_{i=0}^{N} k_i^2}{\sum_{i=0}^{N} k_i} = 10.35886 > 9 = \frac{\sum_{i=0}^{N} k_i}{N}$.

Therefore without full knowledge of a network, rather than choosing a random group of people to monitor, one should ask these random individuals to nominate a friend that they perceived to be more popular (VanderWeele & Christakis 2019). We call this group of people FR and the randomly selected group R, not to be confused with the previous recovered, $R$, group. The randomly selected group R is chosen at every iteration. For example to monitor $n\%$ of the population, we choose randomly choose $n\%$ out of the 148 vertices. In order to obtain the monitored FR group of $n\%$ in our simulations, we randomly select $n\%$ of the population at every iteration and then for each member of this group, we select the individual's most popular friend i.e the friend with the highest degree $k_i$ to be in the $n\%$ FR monitored group as seen in Figures 8 and 9.

The results of simulating various monitoring strategies are displayed in Figures 8 and 9. The values in these figures are listed in Table 1 (See Appendix). In Figure 8, one can clearly see that BC and CC are the significantly better strategies to reduce the total infection in the population and are able to reduce total infection in the population to less than 20%. For monitoring of between 5 and 10% percent of the population, the monitored group based on FR, DC, BC and CC consistently have a lower total infected percentage than the R group.

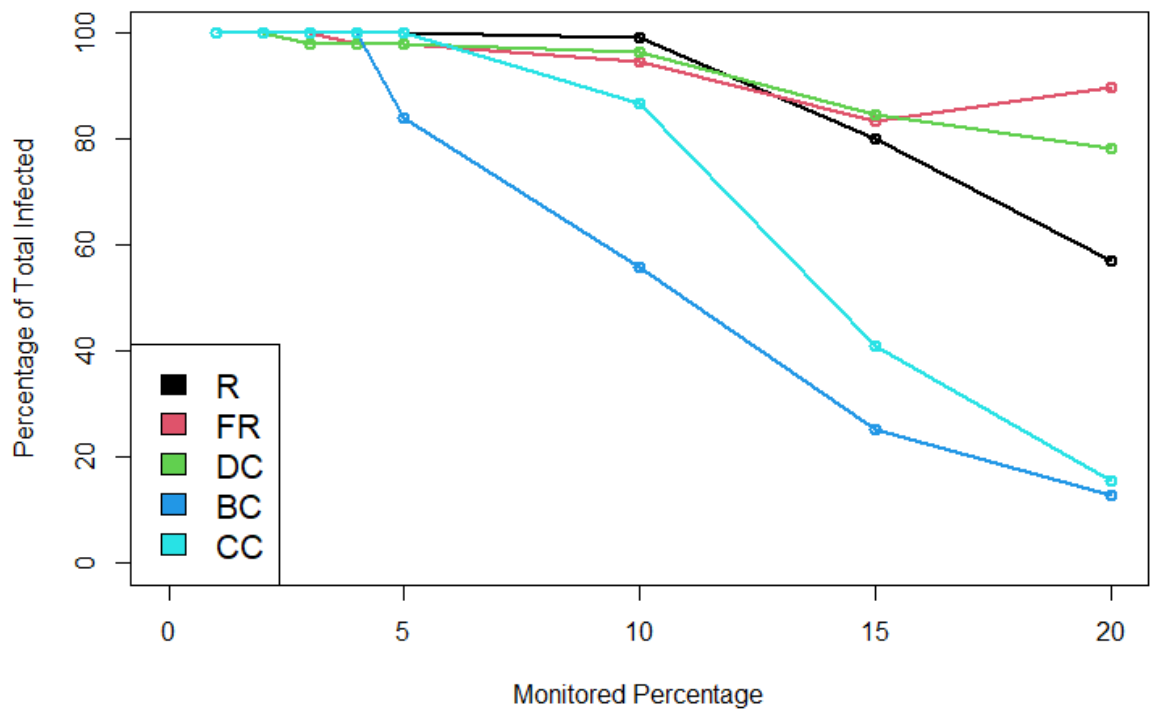

Figure 8. Percentage of total infected individuals from the median of 100 iterations for each monitored group. Colours represent different monitored group.

In Figure 9, we present the percentage of infected individuals at the peak infection time for the five monitoring strategies at 5, 10, 15 and 20% of individuals monitored. The black line representing the randomly chosen monitored group R shows that increasing the monitoring percentage, while reducing significantly the size of the peak only marginally delays it. For the other four groups, not only was the max infected reduced but the peak infection time was also delayed. We also observe that BC and CC strategies are so effective, as seen on Figures 6 and 7, that the peak time is apparently advanced. This is an artefact due to higher monitoring percentage, 15 and 20%, altogether stopping the epidemic early on rather than slowing it down.

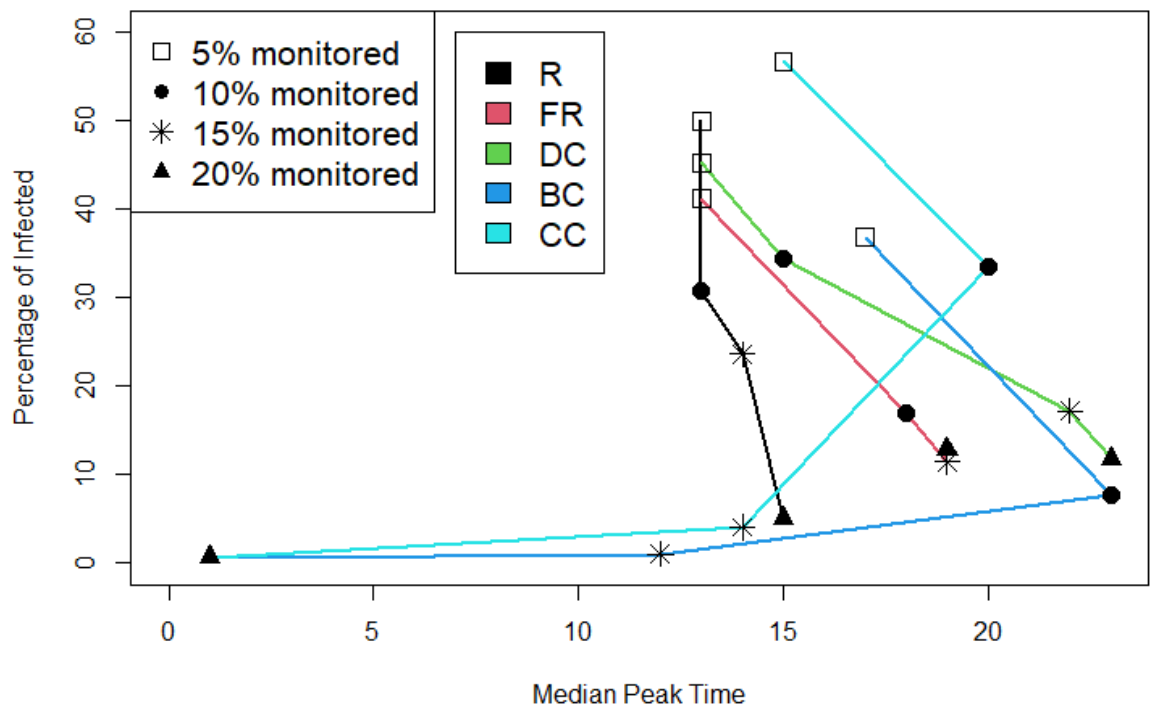

Figure 9. The median peak time and the percentage of the infected at this peak time. Shapes represent different monitored percentage. Colours represent different monitored groups.

From these results, we conclude that given full knowledge of the network, monitoring a group with highest BC and CC will be the most effective in flattening the curve.

**Discussion**

Monitoring and testing certain groups of the population helps to reduce infections. Ideally we need to vaccinate and monitor (test) everyone. However due to the significance of the cost involved, whether in terms of money, time, effort or availability of vaccines, we suggest targeted monitoring strategies that may give more return on investment. A full economic costing is out of the scope of this paper, but hopefully with insights provided by our research, different testing options can be compared in terms of security and cost.

All else being equal, it if comes down to network positions for who to monitor and/or vaccinate, based on the outcome of this study, we recommend prioritising the highest BC group. Otherwise, if knowledge about the network is not available, we recommend prioritising the FR group, with the limitation that a potential bias can be induced if the random selected group R knows about this selection process. The network in this study is a static network i.e. we assume that the connectivity of the network does not change edges over time which introduce two limitations: first we assume that contacts between students will only occur with their reported friends and second considering a static network assumes constant interactions between individuals. In reality, networks are dynamic and if temporal contact information can be collected, via contact tracing apps for example, a more accurate picture can be painted (Lee et al. 2012; Holme & Luis 2019, Cencetti 2021) even when considering approximate static network (Holme 2013). We reiterate our point that contact tracing apps are not accessible for institutions and private companies, so alternative census based methods like the one presented here are still relevant, although alternatives to apps can be considered, e.g. proximity sensor based interaction measurements (Ozella 2012). We point out that while we expect our results to be generalisable to larger network, the drastic effect of the BC and CC strategies at high monitoring percentage could also be an effect of the size of the network and it's particular structure.

Moreover, our analysis is focused on friendship relationships in a physical lecture hall of 148 students, implying a protective "bubble" from the outside world. We foresee that in the new norm, reopening universities will come with social distancing rules and "student group bubbles", where students are grouped by lectures, or series of lectures and should not be interacting with students following lectures they do not themselves attend, and this can be planned.

It is likely that clustering students into a single bubble is not practical as different curricula often share lectures. Identifying student bridging bubbles is easy, and our analysis generalise to bubbles interactions networks. Applying the same methodology on these bubble interaction networks might yield a good approximation of the full contact network of the whole institution and provide good monitoring targets. Since betweenness centrality is a good marker for monitoring targets, students with the most transversal curriculum should be prioritised for monitoring, but this can also be influenced by their friends' groups within different bubbles.

A bubble approach is effectively already implemented by some industries and institutions where teams within departments can only attend work of certain prescribed days or hours. We recommend analysing the information collected from these approaches using network analysis. Future research includes further utilizing the friendship paradox to give insight into how to adapt the monitoring strategies to accommodate for non-compliance to bubbles and potential out of university/institution social interactions.

**Conclusion**

In this paper we have for the first time, to the best of our knowledge, modelled the spread of Covid-19 based on a real Malaysian contact network. We simulated different strategies to be implemented in preparation for meetings in the physical realm and the possible resulting epidemic spread in the event of just one random individual being infected. Three basic measure of effectiveness taken into consideration is the percentage of total infected individuals, the peak infection time and the percentage of infected individuals at peak time, to avoid overloading the health system. The percentage of total infected individuals in Figure 8 and the values at peak infection time in Figure 9, clearly indicates that given full knowledge of the network, monitoring the group with high BC individuals is the best strategy. Without full knowledge of the network, monitoring the recommended friends is a good alternative for small monitoring percentages.

**Acknowledgements**

We are grateful for the grant FRGS/1/2020/STG06/UKM/02/8 for funding our research. Paul Expert is supported by NIHR Imperial BRC (grant number NIHR-BRC-P68711). Many thanks to Jeyasree A/P Ratanarajah for data collection.

**Appendix**

Table 1. The percentage of total infected, maximum infection time and the percentage of infected at maximum time for the median infected curves.

| Group monitored and percentage | Percentage of Total Infected | Max Infection Time | Percentage of Infected at Max Time |
| --- | --- | --- | --- |
| R0 | 100.0 | 12.0 | 84.1 |
| R1 | 100.0 | 13.0 | 72.6 |
| R2 | 100.0 | 13.0 | 63.5 |

| R3   | 100.0 | 13.0 | 62.5 |
| ---- | ----- | ---- | ---- |
| R4   | 100.0 | 13.0 | 60.8 |
| R5   | 100.0 | 13.0 | 50.0 |
| R10  | 99.3  | 13.0 | 30.7 |
| R15  | 80.1  | 14.0 | 23.6 |
| R20  | 57.1  | 15.0 | 5.1  |
| FR1  | 100.0 | 13.0 | 66.9 |
| FR2  | 100.0 | 13.0 | 60.1 |
| FR3  | 100.0 | 12.0 | 52.0 |
| FR4  | 98.0  | 14.0 | 56.4 |
| FR5  | 98.0  | 13.0 | 41.2 |
| FR10 | 94.6  | 18.0 | 16.9 |
| FR15 | 83.4  | 19.0 | 11.5 |
| FR20 | 89.9  | 19.0 | 12.8 |
| DC1  | 100.0 | 12.0 | 69.3 |
| DC2  | 100.0 | 12.0 | 71.3 |
| DC3  | 98.0  | 13.0 | 52.0 |
| DC4  | 98.0  | 14.0 | 44.6 |
| DC5  | 98.0  | 13.0 | 45.3 |
| DC10 | 96.6  | 15.0 | 34.5 |
| DC15 | 84.8  | 22.0 | 17.2 |
| DC20 | 78.4  | 23.0 | 11.8 |
| BC1  | 100.0 | 12.0 | 62.2 |
| BC2  | 100.0 | 12.0 | 54.7 |
| BC3  | 100.0 | 15.0 | 52.7 |
| BC4  | 100.0 | 15.0 | 50.7 |
| BC5  | 84.1  | 17.0 | 36.8 |
| BC10 | 55.7  | 23.0 | 7.8  |
| BC15 | 25.3  | 12.0 | 1.0  |
| BC20 | 12.8 | 1.0 | 0.7 |
| CC1 | 100.0 | 12.0 | 72.3 |
| CC2 | 100.0 | 13.0 | 68.2 |
| CC3 | 100.0 | 15.0 | 59.8 |
| CC4 | 100.0 | 14.0 | 59.1 |
| CC5 | 100.0 | 15.0 | 56.8 |
| CC10 | 86.8 | 20.0 | 33.4 |
| CC15 | 40.9 | 14.0 | 4.1 |
| CC20 | 15.5 | 1.0 | 0.7 |